\documentclass[aps,twocolumn,showpacs]{revtex4}
\usepackage{graphicx}
\begin{document}
\title{
Comparison of Nuclear Suppression Effects on Meson Production
at High $p_T$ and $p_L$}
\author{Rudolph C. Hwa$^1$ and  C.\ B.\ Yang$^{2}$}
\affiliation{
$^1$Institute of Theoretical Science and Department of Physics\\
University of Oregon, Eugene, OR 97403-5203, USA\\
$^2$Institute of Particle Physics, Hua-Zhong Normal University,
Wuhan 430079, P.\ R.\ China}
\date{April 2003}

\begin{abstract}
The medium effect on the pion distribution at high $p_T$ in
$AA$ collisions is compared to that of the pion distribution at high
$p_L$ in $pA$ collisions. Both the suppression of the spectra and
the energy losses of the measured pions are studied. Although the
medium effect on $p_T$ is larger than on $p_L$, the difference is
found surprisingly to be not as big as one would naively expect.

\end{abstract}
\maketitle

\section{Introduction}

When partons traverse nuclear medium, whether dense or not,
they lose momenta by scattering and gluon radiation.  At high
transverse momentum $p_T$ the effect can be calculated in
perturbative QCD, although its reliability is not expected for
$p_T < 10$ GeV/c.  At low $p_T$, but high longitudinal
momentum $p_L$, the effect cannot be calculated in pQCD,
although it is known that partons suffer momenta losses also
in the beam direction.  Experimentally, it is the momenta of the
produced hadrons that are measured. How they are related to
the underlying parton $p_T$ and $p_L$ distributions is still
controversial.  But even at the phenomenological level it is
unknown what the relationship is between the properties of
momentum degradation in the transverse and longitudinal
directions.  We attempt to shed some light on that
relationship in this paper.

There are currently good data on high-$p_T\, \pi^{0}$
produced at the Relativistic Heavy-Ion Collider (RHIC) up to
$p_T = 8$ GeV/c  at various centralities \cite{dd}.  One can
therefore deduce the suppression factor from the $p_T$
distributions as a function of $p_T$ and centrality.  There are
no comparable data on the $p_L$ distributions from RHIC.  At
lower energies the data of NA49 at the SPS provide a good
description of the effect of baryon stopping in $pA$ collision
\cite{hf}.  What we need for comparison with the $p_T$ of
$\pi^{0}$ is the $p_L$ distribution of the produced pions,
for which no data are yet available.  However, we do have a
model calculation of the pion distribution that contains the
degradation effect extracted from the observed proton distribution
\cite{hy}; that will be our input in our study of the suppression
factor in the
$p_L$ distribution.

The two features that we shall compare are very different:  the
degradation of $p_T$ in $Au + Au$ collisions at RHIC and the
degradation of $p_L$ in $p + Pb$ collisions at SPS.  In the
absence of any information in the literature on the quantitative
or qualitative difference between the two types of degradation
effects of the nuclear medium, even a crude estimate of the
suppression properties described in a common language
would be illuminating.   Without a study of the type proposed
here, one does not even know whether the strengths of
suppression are within the same order of magnitude,
especially since the medium in $AA$ collision at RHIC is
dense and hot, while the medium in $pA$ collision at SPS is
uncompressed and cold.

There is a theoretical issue that is of interest to discuss here.
For some time there has been a school of thought that all
hadrons are produced by the fragmentations of partons, not
only in the transverse direction in the form of jets (which is
generally accepted), but also in the longitudinal direction in
the form of breaking of strings (as in the dual parton model
\cite{dpm}).  The effect of the nuclear medium has been
represented by the modification of the fragmentation function
\cite{wz} in transverse direction, but that of the longitudinal
direction is not known.  However, fragmentation of partons is
not the only way to produce hadrons.  Recent investigations have
shown that quark recombination can be important in the high
$p_T$ problem \cite{hy2}-\cite{fm}, in addition to its relevance
originally proposed for the high $p_L$ problem \cite{dh,hy3}.
Since the multiparton distributions needed for recombination are
drastically different in the transverse and longitudinal directions,
the effect of the nuclear medium is considerably more
complicated.  In this paper we can avoid dealing directly with those
complications by putting the emphasis on the phenomenology of
the hadrons produced.  For the $p_T$ problem we shall use the
scaling form of the data \cite{hy4}, while for the $p_L$ problem
the calculated $x_F$ distributions that follow from the $pA$
data will be used.  Our task is made easier by not deriving the
$p_T$ and
$p_L$ distributions, but by focusing on the centrality dependences
of those distributions.

Suppression of the meson distribution is like jet quenching at
the parton level. In addition to the study of suppression,
we shall also consider energy loss, which is another way of
quantifying the medium effect. It corresponds to a shift in
$p_T$ or $p_L$ that is necessary for the inclusive cross
section in medium to be equivalent to a reference cross section
with minimal medium effect. We shall find interesting results
in the shift that are very different from the prediction of
pQCD. That difference is much larger than the difference
between the effects on the transverse and longitudinal motions
of the produced mesons.

Since our comparison is between $p_T$ in $AA$ collisions and
$p_L$ in $pA$ collisions, they are two steps removed from
each other.  When good $p_L$ data on identified pions in $AA$
collisions become available, they shall then serve as the
intermediate station to make possible two one-step comparisons:
(1) between $p_T$ and $p_L$ in $AA$ collisions, and (2) between
$p_L$ in $AA$ and $p_L$ in $pA$ collisions.  What we do here
therefore sets the stage for that work to come.

\section{Suppression of the Pion $p_T$ Distribution in $AA$
Collisions}

For the $p_T$ distribution of pions in nuclear collisions we
use the PHENIX data on $\pi^{0}$ production at
midrapidity with $p_T$ extending to as high as 8 GeV/c
\cite{dd}.  A convenient scaling form for that distribution has
been found that summarizes the dependence on energy and
centrality in terms of a simple analytical formula
\cite{hy4,hy5}.  The quantity that we want to study is the
suppression factor $S$, which is the ratio of the normalized
$p_T$ distribution $P$:
\begin{eqnarray}
S (x_T, N) = P(x_T, N)/P(x_T, 2) \ ,
\label{1}
\end{eqnarray}
where $x_T$ is the scaled $p_T$ variable
\begin{eqnarray}
x_T = p_T/K_0 \ ,
\label{2}
\end{eqnarray}
and $N$ is the abbreviated notation for the number of
participants $N_{\rm part}$.  The scale $K_0$ is set at 10
GeV/c for convenience; it is trivial to move it higher when
higher
$p_T$ data become available.  If $x^{-1}_T d N_{\pi}/dx_T$
denotes the $x_T$ distribution of produced $\pi^{0}$,
averaged over midrapidity and over all azimuthal angle
$\phi$, then $P(x_T, N)$ is defined by
\begin{eqnarray}
P(x_T, N) = {1 \over x_T}{d N_{\pi}  \over dx_T
} \,\left/ \int^1_0dx_T{d N_{\pi} \over  dx_T}\right. \ .
\label{3}
\end{eqnarray}

Instead of determining $P(x_T, N)$ directly from the data for
every $N$ at any given $s$, it is simpler to derive it from the
scaling function $\Phi(z)$ that is an excellent fit of all
high-energy data for all centralities \cite{hy4,hy5}.  Since
the details of the relationship between $P(x_T, N)$ and
$\Phi(z)$  are given in Refs.\ \cite{hy4,hy5}, there is no need
for us to repeat them here.  For the convenience of the reader,
they are summarized in the Appendix.

\begin{figure}[tbph]
\includegraphics[width=0.45\textwidth]{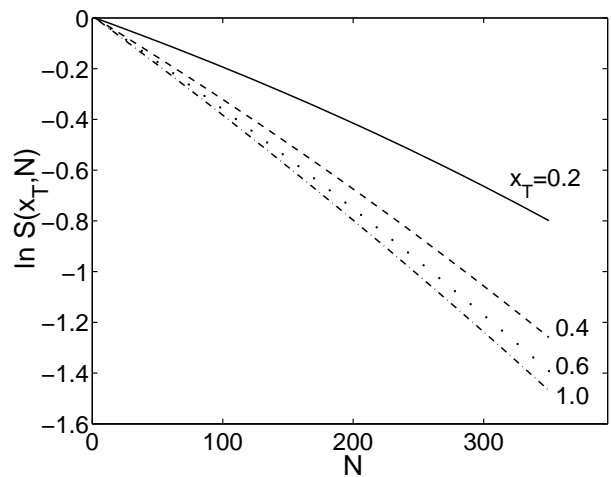}
\caption{The dependences of the suppression factor $S(x_T, N)$
on the number of participants $N$ for various values of $x_T$.}
\end{figure}

  From the analytical expression for $S(x_T, N)$ obtained by
use of Eq. (\ref{A.15}) in Eq. (\ref{1}) we can examine the $N$
dependence for fixed values of $x_T$ by plotting ${\rm ln}
S (x_T, N)$ vs $N$ for some sample values of $x_T$, as shown
in Fig.\ 1.  Approximating the nearly linear behaviors in Fig.\ 1
by straight lines, we obtain
\begin{eqnarray}
{\rm ln} S(x_T, N) = a_0 (x_T) - a_1 (x_T)N \ ,
\label{4}
\end{eqnarray}
where $a_0 (x_T)$ and $a_1 (x_T)$ are shown in Fig.\ 2.
The solid lines in that figure are fits, using the parametrization
\begin{eqnarray}
a_0 (x_T) = 0.031 \left[ 1 - \exp (0.5 - 12 x_T)\right] \ ,
\label{5}
\end{eqnarray}
\begin{eqnarray}
a_1 (x_T)= 0.0043 \left[ 1 - \exp (0.2 - 5 x_T)\right] \ .
\label{6}
\end{eqnarray}
A general statement that can be made about ${\rm ln}  S (x_T,
N)$ is that it is approximately linear in $N$, and that the
parameters of the linear fits are roughly independent of $x_T$
when $x_T \geq 0.4$.

\begin{figure}[tbph]
\includegraphics[width=0.45\textwidth]{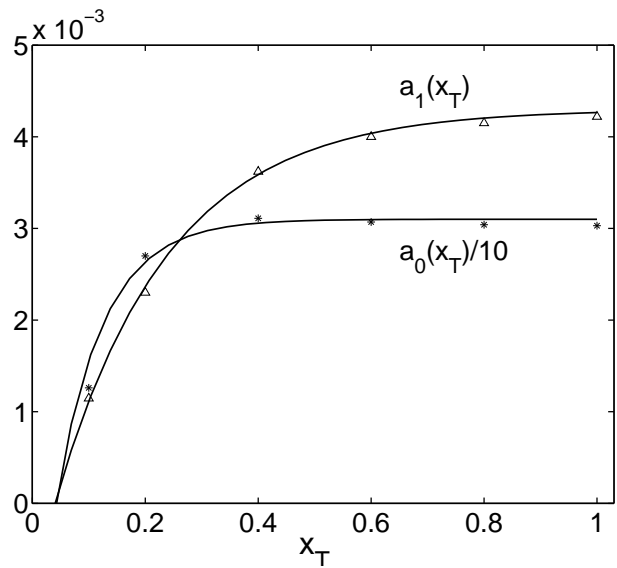}
\caption{The coefficients $a_0(x_T)$ and $a_1(x_T)$, where
$a_0(x_T)$ is plotted with reference to the scale on the left, while
$a_1(x_T)$ refers to the scale on the right.}
\end{figure}

Since the number of participants $N$ is associated only with
$AA$ collisions, we must express it in terms of some measure of
nuclear length in order to be able to compare the suppression
factors in $AA$ and $pA$ collisions.  The dependence of $N$
on the impact parameter $b$ is known \cite{kn}.  At a fixed
$b$, the lens-shaped overlap region in the transverse plane of
two colliding nuclei of the same radius $R_A$ has a minimum
distance between the center of the overlap and the edge of
either nuclei (assumed to have a sharp boundary)
\begin{eqnarray}
L_{\rm min}  = R_A - b/2 \ .
\label{7}
\end{eqnarray}
The maximum distance between the center of the overlap to
the edge of both nuclei is
\begin{eqnarray}
L_{\rm max}  = \left(R^2_A - b^2/4 \right)^{1/2} \ .
\label{8}
\end{eqnarray}
The distance that a parton would travel in the nuclear medium
at midrapidity in the transverse plane can be as large as
$2L_{\rm max}$ and as small as 0, depending on where the
parton starts and in which direction.  We shall set the average
distance $L$ traversed to be
\begin{eqnarray}
L(b) = L_{\rm min}  = R_A - b/2 \ ,
\label{9}
\end{eqnarray}
which is an approximate average over all azimuthal angles and
origins of parton paths. Any more detailed geometrical
averaging is pointless, since the nuclear density of the overlap
region is not uniform, and the nonuniformity depends on
$b$.  Knowing $N(b)$ and $L(b)$ enables us to plot $N$ vs
$L$, as shown by the solid line in Fig.\ 3.  We shall
approximate $N(L)$ by a straight line
\begin{eqnarray}
N(L) = - 35.8 + 62.2 L
\label{10}
\end{eqnarray}
(with $L$ in units of fm), shown by the dashed line in Fig.\ 3.  In
view of the nonuniformity of the nuclear density traversed by a
parton in the overlap regions, we believe that an approximation of
$N(L)$ by a linear dependence in Eq.\ (\ref{10}) is good
enough to represent the path length that enters into the
description of the momentum degradation effect.

\begin{figure}[tbph]
\includegraphics[width=0.45\textwidth]{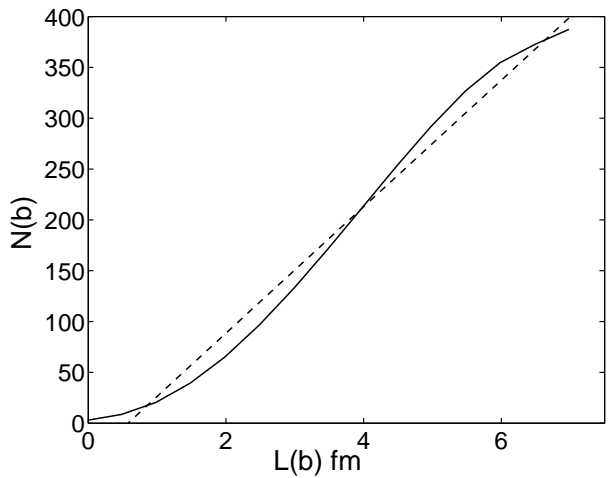}
\caption{The parametric dependence of $N(b)$ on $L(b)$, where
the solid line follows from the use of Eq.\ (\ref{9}) for $L(b)$,
while the dashed line is a straightline fit of the solid line.}
\end{figure}

Substituting Eq.\ (\ref{10}) into Eq.\ (\ref{4}) we obtain
\begin{eqnarray}
{\rm ln} S (x_T, L) = \Lambda_0 (x_T) - \Lambda (x_T)L\ ,
\label{11}
\end{eqnarray}
where in $\Lambda_0 (x_T)$ the contribution from the
$a_0(x_T)$ term in Eq.\ (\ref{4}) is negligible.  Thus Eq.\
(\ref{11}) may be rewritten as
\begin{eqnarray}
{\rm ln} S (x_T, L) = -\Lambda (x_T) (L - 0.71)\ ,
\label{12}
\end{eqnarray}
where
\begin{eqnarray}
\Lambda (x_T) = 0.27 \left[1 - \exp (0.2 - 5 x_T) \right] \ ,
\label{13}
\end{eqnarray}
which is nearly constant for $x_T > 0.5$, as shown by the solid line
in Fig.\ 4. This exponential dependence of $S (x_T, L)$ on $L$ is in a
familiar form for nuclear attenuation.  We remark that
although the length $L$ is the approximate average distance in
the nuclear medium that a parton traverses, no parton
dynamics has been assumed in the derivation of $S (x_T, L)$,
which is extracted from the data on $\pi^{0}$ production
without dynamical modeling.  The exponential form of $S (x_T,
L)$ is similar to the Gerschel-H\"{u}fner formula for the
$J/\psi$ suppression \cite{gh}, except that the latter refers to
a quantity integrated over all $p_T$, and is related to the
dissociation of $J/\psi$ in the medium.  The quarks and
antiquarks that form the produced $\pi^{0}$ in our
expression for $S (x_T, L)$ are the results of gluon radiation
and gluon conversion processes, most of which are not
calculable in pQCD, especially near the end of the evolution
process.

\begin{figure}[tbph]
\includegraphics[width=0.45\textwidth]{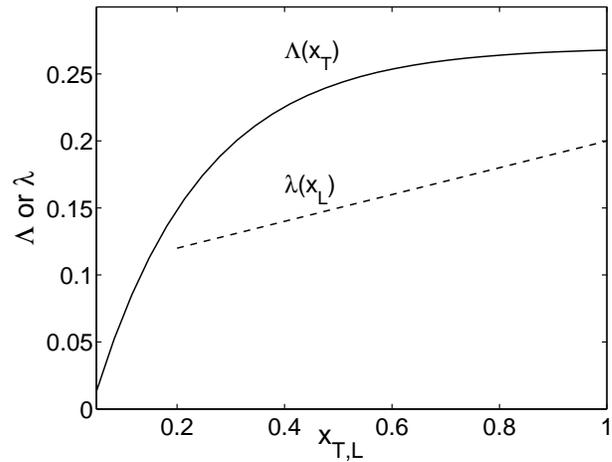}
\caption{ The solid line is for $\Lambda(x_T)$ and the dashed line
is for $\lambda(x_L)$. The horizontal axis can be for either $x_T$
or $x_L$.}
\end{figure}

\section{Suppression of the Pion $p_L$ Distribution in $pA$
Collisions}

Since no $p_L$ distributions of identified particles in the
fragmentation regions of $AA$ collisions at RHIC are available,
we consider the processes at the SPS energies.  At lower
energies the projectile fragmentation region can contain
particles arising from the fragmentation of the target, and
vice-versa.  It is therefore important to consider reactions that
are free of such `spill-over' particles.  The preliminary data of
NA49 on $pA$ collisions at SPS have been analyzed to provide
the produced $p-\bar{p}$ distribution of the projectile
fragmentation only \cite{hf}.  That is accomplished by
determining the $p_L$ distribution
\begin{equation}
F_p (p-\bar{p}) = (p-\bar{p})_p - {1 \over 2}\left[
(p-\bar{p})_{\pi ^{+}} + (p-\bar{p})_{\pi ^{-}}\right] \ ,
\label{15}
\end{equation}
where $(p-\bar{p})_h$ denotes the distribution of $h + A
\rightarrow (p-\bar{p}) + X$, since the quantity inside the
square brackets contains no beam fragments by charge
conjugation symmetry.  A similar distribution for pions, $F_p
(\pi^+ - \pi^-)$, unfortunately does not exist, which is what
we need for comparison with the result obtained in the
preceding section.

Although there are no experimental data for pion production
in the proton fragmentation region free of target fragments,
theoretical results on such distributions are available that are
based on the centrality dependence of the experimental data
of $F_p (p-\bar{p})$.  In Ref. \cite{hy} the recombination
model is used to relate the data to the nuclear degradation
effect on the $p_L$ of the produced $p-\bar{p}$, which in turn
is then used to predict the $p_L$ distributions of the produced
pions.  Since the nuclear suppression factor on the produced
pions is directly related to the experimental suppression of the
produced $p-\bar{p}$, we shall use the result in \cite{hy} for
the pion distribution in $p + Pb$ collisions.

Let us use $H (x_L,\bar{\nu})$ to denote the pion inclusive
distribution in the scaled longitudinal momentum, integrated
over $p_T$, i.e.,
\begin{equation}
H (x_L,\bar{\nu}) = x_L {dN_{\pi} \over dx_L}\ ,
\label{16}
\end{equation}
where $x_L = 2 p_L/\sqrt{s}$ and $\bar{\nu}$ is the average
number of collisions with nucleons in the target nucleus.
  From Fig.\ 9 in Ref. \cite{hy} one sees that $H
(x_L,\bar{\nu})$ is essentially exponential in $x_L$ for
various values of $\bar{\nu}$.  We use the following
parameterization for $p + Pb \rightarrow \pi^+ + X$
\begin{equation}
H (x_L,\bar{\nu}) = \exp \left[- h_0 (\bar{\nu}) - h_1
(\bar{\nu}) x_L \right] \ ,
\label{17}
\end{equation}
where
\begin{eqnarray}
h_0 (3.1) = 0.92,&  &h_0 (6.3) = 1.61\ , \nonumber\\
h_1 (3.1) = 5.89,& & h_1 (6.3) = 6.57\ .
\label{18}
\end{eqnarray}
These parameters provide a good fit for $0.2 \leq x_L \leq
0.6$.  For $x_L > 0.6$, the slopes are slightly higher, but the
ratio of $H (x_L,\bar{\nu})$ at the two values of $\bar{\nu}$
is about the same.

Since the data are for the two values of $\bar{\nu}$ only, our
definition of the suppression factor is
\begin{equation}
S\left(x_L; \bar{\nu}_2, \bar{\nu}_1 \right) =
H\left(x_L,\bar{\nu}_2\right) \,\left/
H\left(x_L,\bar{\nu}_1\right) \right.
\label{19}
\end{equation}
with $\bar{\nu}_1 = 3.1$ and $\bar{\nu}_2 = 6.3$.  We now
make the assumption that $h_0 (\bar{\nu})$ and $h_1
(\bar{\nu})$ are linear functions of $\bar{\nu}$, partly
because the average $\langle x_L \rangle$ can be
shown to decrease exponentially with $\bar\nu$ \cite{hy}, and
partly because all nuclear damping effects are empirically
dependent on the path length in exponential form.  With that
assumption we can express
$S\left(x_L;
\bar{\nu}, 1\right)$ for any $\bar{\nu}$ relative to
$\bar{\nu} = 1$ as
\begin{equation}
{\rm ln} S\left(x_L;  \bar{\nu}, 1\right) = -
\left[\theta_0 (\bar{\nu}) + \theta_1 (\bar{\nu}) x_L \right]
\label{20}
\end{equation}
where $\theta_0 (\bar{\nu})$ and $\theta_1 (\bar{\nu})$ are
the linear functions of $\bar{\nu}$
\begin{equation}
\theta_i (\bar{\nu}) = h_i(\bar{\nu})-h_i (1) =
\theta^{\prime}_i \, (\bar{\nu} - 1) \ , \qquad i=0,1\ .
\label{21}
\end{equation}
The slopes are
\begin{equation}
\theta^{\prime}_i = {h_i (\bar{\nu}_2) - h_i (\bar{\nu}_1)
\over  \bar{\nu}_2 - \bar{\nu}_1} \ .
\label{23}
\end{equation}
  From the values given in Eq.\ (\ref{18}), we find that
$\theta^{\prime}_0$ and $\theta^{\prime}_1$ are very
nearly equal.  We denote them collectively by
\begin{equation}
\theta^{\prime} = 0.21
\label{24}
\end{equation}
and Eq.\ (\ref{20}) becomes
\begin{equation}
{\rm ln} S\left(x_L;  \bar{\nu}, 1\right) = -
\theta^{\prime} (1 + x_L) (\bar{\nu} - 1) \ .
\label{25}
\end{equation}
The exponential dependence of $S$ on $\bar{\nu}$
   is now explicit.  The specific $x_L$ dependence of the decay
coefficient is a result of the distribution in Eq.\ (\ref{17})
calculated in \cite{hy}.

If $L$ denotes the path length in a $pA$ collision, then the
usual expression for $\bar{\nu}$ in terms of $L$ is
\begin{equation}
\bar{\nu} = \sigma_{pp} \rho L, \qquad \rho = {4\pi
\over 3} {A \over R^3_A}\ .
\label{26}
\end{equation}
Using $R_A = 1.2 A^{1/3}$ fm, and $\sigma_{pp}  = 35$ mb,
we have
\begin{equation}
\bar{\nu} = 0.48 L \ ,
\label{27}
\end{equation}
where $L$ is in units of fm.  Substituting this in Eq.\ (\ref{25})
yields
\begin{equation}
{\rm ln} S\left(x_L,  L\right) = -
\lambda (x_L)\ (L-2.1) \ ,
\label{28}
\end{equation}
where
\begin{equation}
\lambda (x_L) = 0.1\ (1 + x_L) \ .
\label{29}
\end{equation}

Equations (\ref{28}) and (\ref{29}) for the suppression of the
$p_L$ distribution in $pA$ collisions are the counterparts of
Eqs.\  (\ref{12}) and (\ref{13})  for the suppression of the
$p_T$ distribution in $AA$ collisions.  Instead of $\Lambda
(x_T)$ that rises rapidly at small $x_T$, but saturates to
a nearly constant value for $x_T > 0.5$, as shown in Fig.\ 4, we
now have $\lambda (x_L)$ which is linearly rising, shown by the
dashed line in Fig.\ 4.  Since Eq.\ (\ref{17}) is not reliable for
$x_L < 0.2$, we do not show that portion of $\lambda (x_L)$ in
Fig.\ 4. For numerical comparison we can consider
$x_T = x_L$ at two values, 0.6 and 0.8:
\begin{eqnarray}
\Lambda(0.6) = 0.25 \ , \quad\quad  \lambda (0.6) = 0.16
\ ,
\nonumber\\
\Lambda(0.8) = 0.26 \ , \quad\quad  \lambda (0.8) = 0.18
\ .
\label{30}
\end{eqnarray}
Evidently, $\Lambda (x_T)$ is larger than $\lambda (x_L)$
when $x_T = x_L > 0.5$, but not by much.  Of course, there is
no cogent reason to compare them at $x_T \approx x_L$ since
the scale $K_0$ in the definition of $x_T$ is arbitrary, while
$x_L$ is Feynman $x_F$ and has scaling property.  However, with
$\Lambda (x_T)$ being roughly constant for $x_T > 0.5$, it
does not matter what $x_T$ is exactly.  The relative values of
$\Lambda (x_T)$ and $\lambda (x_L)$ shown in Eq.\
(\ref{30}) give us some indication of how they differ.

\section{Energy Losses in $p_T$ and $p_L$}

Another way to quantify the medium effect is in terms of energy
loss. Since our aim in this paper is to stay at the level of
observable quantities, we cannot descend to the parton level where
the concept of energy loss makes sense as one can study the
evolution of a parton in its trajectory through the nuclear or quark medium.
A produced meson does not itself traverse that medium, since it is
only at the end of the evolution of the parton system that it is formed.
Nevertheless, the notion of energy loss can be expressed in terms
of a shift in $x_T$ in comparing the meson inclusive distributions
at two different values of $N$. To be specific, let us set the
reference value of $N$ at $N=2$, which is not exactly $pp$
collision, but is low enough to represent minimal nuclear effect.
Let us then define the shift $X$ in $x_T$ by
\begin{equation}
P(x_T, N)=P(x_T+X, 2) \ ,   \label{4.1}
\end{equation}
where $P(x_T, N)$ is the normalized pion distribution defined in
Eq.\ (\ref{3}). Thus $X$ measures the degradation of the pion
$x_T$ in changing $N_{\rm part}$ from 2 to $N$.

Using Eq.\ (\ref{A.15}) where $\left<x_T\right>$ is calculable
by use of Eq.\ (\ref{A.13}), we can solve Eq.\ (\ref{4.1}) and
determine $X$ in terms of $x_T$ and $N$. At fixed $N$, the
dependence of $X$ on $x_T$ is nearly perfectly linear.
Parametrisizing it as
\begin{equation}
X(x_T, N)=x_0(N) + \xi(N)\ x_T \ ,    \label{4.2}
\end{equation}
we find, for $N=100, 200,$ and 350, the result $x_0(N)=-0.0047,
-0.0023,$ and $-0.0039$, and $\xi(N)=0.0432, 0.0916,$ and 0.1741,
respectively. Since $x_0$ is negligible except at very small
$x_T$, we can regard $\xi(N)$ as the fractional shift, or more
precisely as
\begin{equation}
\xi(N)=dX(x_T, N) / dx_T \ ,    \label{4.3}
\end{equation}
whose dependence on $N$ is shown in Fig.\ 5.  Disregarding the
point at $N=2$, where by definition $\xi(2)=0$, the three points
at $N=100, 200,$ and 350 can be fitted by a straight line, as
shown by the solid line in Fig.\ 5. Its parametrization is
\begin{equation}
\xi(N)=-0.011+5.26\times 10^{-4}\ N        \label{4.4}
\end{equation}
for $N\stackrel{>}{\sim} 50$. Thus for $AA$ collisions the
fractional energy loss or fractional shift in $p_T$ is
independent of $p_T$ and depends linearly on $N$ with a
coefficient
\begin{equation}
\xi'_N=5.26\times 10^{-4} \ .      \label{4.5}
\end{equation}

\begin{figure}[tbph]
\includegraphics[width=0.45\textwidth]{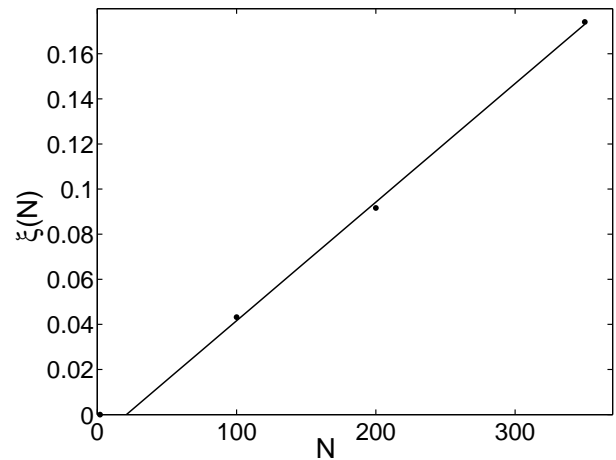}
\caption{ The $N$ dependence of the fractional shift $\xi$. The
solid line is a linear fit of the upper three points.}
\end{figure}

This result can be obtained quickly, but only approximately, in
two ways. First, if one ignores the dependence on $N$ of the
denominator in Eq.\ (\ref{A.8}), an approximation that amounts to
setting $k(s,N)$ to be a constant (an error $\le 20\%$) compared
to the many orders of magnitude of variation of $\Phi(z)$, then
the condition of Eq.\ (\ref{4.1}) is equivalent to identifying its
two sides with the same function $\Phi(z)$ with $z$ evaluated at
$N_{\rm part}=N$ and 2, i.e.,
\begin{eqnarray}
\xi(N)&\approx& {p_T(2)-p_T(N)\over p_T(N)}={K(2)-K(N)\over
K(N)}\nonumber\\
&&={6.36\times10^{-4}\ (N-2)\over K(N)} \ .  \label{4.6}
\end{eqnarray}
Setting $K(N)\approx K(2)=1.226$ results in
\begin{equation}
\xi'_N\approx 5.19\times 10^{-4} \ ,       \label{4.7}
\end{equation}
which is only slightly less than the more accurate value in Eq.\
(\ref{4.5}).  The other approximate way of determining the
fractional shift is to use Eq.\ (\ref{A.15}) and ignore the $N$
dependence of $\left<x_T\right>^2$ in that equation so that the
condition in Eq.\ (\ref{4.1}) becomes an identification of the $u$
variable in $\Psi(u)$ at two values of $N$, i.e.,
\begin{equation}
u(N)={x_T\over \left<x_T\right>_N}=u(2)={x_T+X\over
\left<x_T\right>_{N=2}} \ .    \label{4.8}
\end{equation}
Since $\left<x_T\right>_N$ depends on $N$ in a known way \cite{hy4}
\begin{equation}
\left<x_T\right>_N=\left<x_T\right>_{N_0}\ \exp
[\,-\beta\,(N-N_0)] \ ,   \label{4.9}
\end{equation}
where $\beta=5.542\times 10^{-4}$ (denoted by $\lambda$ in
\cite{hy4}), its use in Eq.\ (\ref{4.8}) results in
\begin{equation}
\xi'_N\approx \beta=5.54\times10^{-4} \ .    \label{4.10}
\end{equation}
Note that the smallness of $\beta$ in Eq.\ (\ref{4.9}) roughly
justifies the treatment of $\left<x_T\right>$ as a constant in
Eq.\ (\ref{A.15}) in the first place. The value of $\xi'_N$ in
Eq.\ (\ref{4.10}) is only slightly larger than that in Eq.\
(\ref{4.5}). Thus the two approximate methods yield results that
bracket the correct value closely, and illustrate the crucial
role that the scaling variables $z$ and $u$ play.

It is of interest to note that in pQCD the shift in $p_T$ of the
vacuum spectrum necessary to effect the in-medium spectrum is
proportional to $\sqrt p_T$ \cite{bdm}. It means that the
fractional shift decreases with $p_T$, whereas our
phenomenological result indicates that it is independent of $p_T$.
However, since pQCD is reliable only for $p_T>10$ Gev/c, while our
analysis is based on data at $p_T<10$ Gev/c, there is as yet no
direct conflict. Nevertheless, the disagreement in the $p_T$
dependences is worth bearing in mind.

We now consider the energy loss of $p_L$ of pions produced in $pA$
collisions. We define the shift $X$ by referring the inclusive
cross section at $\bar\nu$ to that at $\bar\nu=1$, i.e.,
\begin{equation}
H(x_L, \bar\nu)=H(x_L+X, 1) \ .    \label{4.11}
\end{equation}
The use of Eq.\ (\ref{17}) and (\ref{21}) leads to
\begin{equation}
X(x_L, \bar\nu)=\xi(\bar\nu)\ (x_L+1) \ ,    \label{4.12}
\end{equation}
where
\begin{equation}
\xi(\bar\nu)=\xi'_{\bar\nu}\ (\bar\nu-1) \ ,    \label{4.13}
\end{equation}
\begin{equation}
\xi'_{\bar\nu}=\theta'/h_1(1)=0.0385 \ .     \label{4.14}
\end{equation}

To compare with the result from $AA$ collisions, let us convert
both $N$ and $\bar\nu$ to the average path length $L$. Using Eq.\
(\ref{10}) in (\ref{4.4}), we get
\begin{equation}
\xi_T(L)=-0.03+0.0327\ L \ .    \label{4.15}
\end{equation}
Using Eq.\ (\ref{27}) in (\ref{4.13}), we get
\begin{equation}
\xi_L(L)=-0.0385+0.0185\ L \ .    \label{4.16}
\end{equation}
In both cases the fractional shifts depend linearly on $L$ with
the coefficients
\begin{equation}
\xi'_T=0.0327\ , \qquad\qquad \xi'_L=0.0185 \ .    \label{4.17}
\end{equation}
The ratio of these two coefficients is very nearly the same as the
ratio of $\Lambda(x_T)$ to $\lambda(x_L)$ in the region
$x_{T,L}\approx0.5$ [cf. Fig.\ 4].
Thus the study of energy loss and that of suppression give
comparable results on the effects of the nuclear medium. The
advantage of using Eq.\ (\ref{4.17}) for comparing the nuclear
effects on the transverse and longitudinal motions is that the
fractional shift is independent of the scale $K_0$ used in the
definition of $x_T$ in Eq.\ (\ref{2}). Thus the numerical values
of $\xi'_N, \xi'_{\bar\nu}, \xi'_T$ and $\xi'_L$ are simple
quantitative results of this investigation that can be
reexamined in the future when experimental data at different
energies for different colliding nuclei become available.

\section{Conclusion}

We have studied the suppression effect of the medium and the
energy losses of the produced particles. The former is a
comparison of the inclusive distributions in-medium versus
minimal-medium at the same $x_T$ or $x_L$. The latter is the
shift in $x_T$ or $x_L$ necessary for the two distributions at
different $N$ to be equivalent. Both studies yield
qualitatively the same level of effect. In the following we
shall use the suppression effect to represent both in our
discussion of the differences between the transverse and
longitudinal effects.

The primary remark to make is that the values of $\Lambda
(x_T)$ and $\lambda (x_L)$ given in Eq.\ (\ref{30}) are
amazingly close.  There are many arguments one can give to
suggest that $\Lambda (x_T)$ and $\lambda (x_L)$ should not
be similar in magnitude, and few are available to explain that
they are even within the same order of magnitude.  Let us
present some of them.

The main difference between the two suppression effects is
that one refers to transverse, the other longitudinal motion.
For transverse momenta of partons caused by hard collisions,
at least there is pQCD to describe some aspects of the
dynamics, although not reliably for $p_T < 8$ GeV/c.  For the
longitudinal momenta of the particles detected, there is no
basic theory to describe their behavior in terms of partons
without some substantial use of models.  Some properties of
the $p_T$ degradation can be calculated, but a recent result
on the nuclear modification factor does not reproduce even
the trend of the $p_T$ dependence \cite{sw}.  Nevertheless,
much more work has been done in applying pQCD to the high
$p_T$ problem than to the high $p_L$ problem.
Since no large momentum transfers are involved at high
$p_L$, one would naively not expect the nuclear suppression
effect to be of the same nature as at high $p_T$.  Yet we find
$\Lambda (x_T)$ and $\lambda (x_L)$ to be comparable.

The media of the two problems are also different.  In $AA$
collisions at RHIC one expects the nuclear medium to be dense
and hot, if not a quark-gluon plasma.  In $pA$ collisions at SPS
the medium that the projectile traverses is the normal
uncompressed nucleus.  One would therefore expect the effects
of the media on momentum degradation to be very different,
yet they are not.

The estimates of the average path length $L$ involve different
approximations for the two cases, and it is difficult to assess
the effects of those approximations.  The only way that $p_T$ in
$AA$ collisions can be compared to $p_L$ in $pA$ collisions is
in the common language of exponential decay in terms of a
path length $L$.  To improve on this problem,  we have to
gain more information from  experiments at intermediate steps
by measuring
$p_L$ in $AB$ collisions with various nuclear sizes of $A$.

We can think of one reason that could possibly explain the
closeness of $\Lambda (x_T)$ to $\lambda (x_L)$.  The
inclusive cross sections that we examine for the calculation of the
suppression factor $S$ are for pions, not nucleons or other
baryons.  Whereas leading baryons are strongly related to
the valence quarks in the projectile, the pions are more
associated with the gluons, which undergo conversion to
$q\bar{q}$ pair before hadronization.  The depletion of
gluons in the nuclear medium can lead to pion suppression in
any direction.  Evidence for gluon depletion even in $pA$
collisions can be found in the suppression of $J/\psi$
production at large $x_L$ \cite{jp,hpp}.  Our result on the
closeness of $\Lambda (x_T)$ and $\lambda (x_L)$ may well
suggest that gluon depletion is the main mechanism for the
suppression of both $p_T$ and $p_L$ distributions.

In quantitative terms it must be recognized that $\Lambda(x_T)$
is undisputatively larger than $\lambda(x_L)$ and therefore
provides some comfort that the suppression effect is enhanced
when the medium is denser and hotter. What is unexpected is that
it is not an order of magnitude larger.  In order to fully
understand the suppression problem we need  a whole set of
experiments that measure identified pions at high $p_T$ and
$p_L$ for all combinations of nuclear sizes in $AB$ collisions
at high energy.  It is important to discover what is universal in the
nuclear effects on the produced particles, and what is not. The
finding in this paper constitutes an interesting and intriguing
beginning in that direction.

\section*{Acknowledgment}

We acknowledge helpful discussions with H.\ Huang and A.\ Tai at
the early stage of this work.  This work was supported, in part,
by the U.\ S.\ Department of Energy under Grant No.
DE-FG03-96ER40972.

\section*{Appendix}

In this Appendix we summarize the basic formulas that relate
$P(x_T, N)$ to two scaling functions found in \cite{hy4}.  The
first scaling function is
\begin{equation}
\Phi(z)=1200\,(z^2+2)^{-4.8}\left(1+25\,{\rm e}^{-4.5z}\right)\ ,
\label{A.1}
\end{equation}
which describes the $p_T$ distributions at all $N$  (number
of participants) and all $\sqrt{s}$ in terms of
one scaling variable
\begin{equation}
z=p_T/K(s,N)\ ,
\label{A.2}
\end{equation}
where $K(s,N) = K(s) K(N)$, with \cite{hy2,hy5}
\begin{equation}
K(s)=0.69+1.55\times10^{-3}\sqrt{s}\ ,
\label{A.3}
\end{equation}
\begin{equation}
K(N)=1.226-6.36\times10^{-4}N\ ,
\label{A.4}
\end{equation}
$\sqrt{s}$ being in units of GeV and $K(s,N)$ in units of
GeV/c. $\Phi(z)$ is related to the $x_T$ distribution by
\begin{equation}
\Phi(z)=A(N) k^2(s,N) {1\over x_T}{d N_{\pi} \over  d x_T}\ ,
   \label{A.5}
\end{equation}
where rapidity density is implied, and
\begin{equation}
A(N) = 530N_c(N)^{-0.9}\ , \quad
N_c(N) = 0.44N^{1.33}\ ,
\label{A.6}
\end{equation}
and
\begin{equation}
k(s,N) = K(s,N)/K_0 \ ,
\label{A.7}
\end{equation}
$K_0$ being an arbitrary scale, fixed at 10 GeV/c for the
definition of $x_T$ in Eq.\ (\ref{2}).  It is a phenomenological
fact that the combination of separate factors on the RHS of
Eq.\  (\ref{A.5}) that individually depend on $x_T$, $N$ and
$s$ results in a universal function that depends explicitly on
the one variable
$z$ only.

Using Eq.\ (\ref{A.5}) in Eq.\ (\ref{3}) yields
\begin{equation}
P(x_T, N) = \Phi(x_T, N) \left/\int^1_0dx_Tx_T
\Phi(x_T,N)\right.
\label{A.8}
\end{equation}
where $\Phi(z)$ is expressed in terms of $x_T$ and $N$
through
\begin{equation}
z = x_T/k(s,N) \ .
\label{A.9}
\end{equation}
Dependences of $P$ and $\Phi$ on $s$ will not be shown
explicitly.  With Eq.\ (\ref{A.8}), the distribution $P(x_T, N)$
can thus be analytically calculated.  However, instead of
performing the integration in the denominator for every $N$,
there is an even simpler relationship that makes use of
another scaling function.

It is found in Ref.\ \cite{hy4} that there exists another scaling
function
\begin{equation}
\Psi(u)=\Phi(z(u))\,\left/\int du\,u\,\Phi(z(u))\right.,
\label{A.10}
\end{equation}
where
\begin{equation}
u = {p_T  \over  \left<p_T\right>} = {x_T  \over
\langle x_T\rangle} = {z  \over  \langle z\rangle}
   \ .
\label{A.11}
\end{equation}
The new scaling variable $u$ endows $\Psi(u)$ with a property
that is analogous to the Koba-Nielsen-Olesen (KNO) scaling
\cite{kno}.  The average $\langle z\rangle$, defined by
\begin{equation}
\langle z\rangle={\int dz z^2 \Phi(z)\over \int dz z \Phi(z)}
\ ,
\label{A.12}
\end{equation}
is a constant $\left<z\right> = 0.414$, and is related to
$\left<x_T\right>$ by
\begin{equation}
\langle z\rangle = \langle x_T\rangle /k(s,N) \label{A.13}
\end{equation}
due to Eq.\ ({\ref{A.9}}), where $\langle x_T\rangle$ is defined
by
\begin{equation}
\langle x_T\rangle = \int d x_T x_T {dN_{\pi}  \over  d
x_T} \left/\int d x_T  {dN_{\pi}  \over  d
x_T} \right.\ .
\label{A.14}
\end{equation}
A comparison between Eqs.\ (\ref{A.8}) and (\ref{A.10}) yields
\begin{equation}
\Psi(u) = \langle x_T\rangle^2 P (x_T, N)\ .
\label{A.15}
\end{equation}
It is the combination of  Eqs.\ (\ref{A.11}) and (\ref{A.15})
that has led us to regard $\Psi(u)$ as a KNO-type scaling. The
advantage of dealing with $u$ instead of $z$ is that $k(s,N)$
is not explicitly involved in relating $u$ to the observable
$p_T$.  Evaluating Eq.\ (\ref{A.10}), we have
\cite{hy4}
\begin{equation}
\Psi(u)=2.1 \times 10^4 \, (u^2+11.65)^{-4.8} \left(1 +
25e^{-1.864u} \right)  .
\label{A.16}
\end{equation}
   Using Eqs.\ (\ref{A.13}) and (\ref{A.15}), we now have an
algebraic formula for $P(x_T, N)$, which can be used directly
in Eq.\ (\ref{1}) for the suppression factor $S(x_T, N)$.

\end{document}